# Preferential bond formation and interstitial/vacancy annihilation rate drive atomic clustering in gallium ion sputtered compound materials

Zhenyu Ma[1)*], Xin Zhang[1)*], Pu Liu[2)], Yong Deng[1,3)], Wenyu Hu[3)], Longqing Chen[4)], Jun Zhu[5)], , Sen Chen[6)], Zhengshang Wang[7)], Yuechun Shi[8)], Jian Ma[1)], Xiaoyi Wang[1)], Yang Qiu[3)], , Kun Zhang[4)], Xudong Cui[9)], Thomas Walther[10)]

1)Southwest Minzu University, State Ethnic Affairs Commission, Chengdu 610041, China

2)NCS TESTING TECHNOLOGY CO.,LTD, Chengdu 610041, China

3)Pico center, SUSTech Core Research Facilities, Southern University of Science and Technology, Shenzhen 518055, China

4)Key Laboratory of Radiation Physics and Technology of Ministry of Education, Institute of Nuclear Science and Technology, Sichuan University, Chengdu 610064, China

5)College of Physical Science and Technology, Sichuan University, Chengdu 610064, China

6)Laboratory for Shock Wave and Detonation Physics, Institute of Fluid Physics, China Academy of Engineering Physics,Mianyang, Sichuan 621900, China

7)Sichuan Research Center of New Materials, 596 Yinhe Road, Shuangliu, Chengdu 610200, PR China

8)YONGJIANG LABORATORY Ningbo 315000, China

9)Institute of Chemical Materials, China Academy of Engineering Physics, Mianyang 621900, PR China

10)Dept. Electronic & Electrical Eng., University of Sheffield, Mappin St., Sheffield S1 3JD, UK

*Electronic-mail:80300024@swun.edu.cn and qiuy@sustech.edu.cn

(Dated: 8 May 2023)

**Abstract**：The investigation of chemical reactions during the ion irradiation is a frontier for the study of the ion-material interaction. In order to derive the contribution of bond formation to chemistry of ion produced nanoclusters, the valence electron energy loss spectroscopy (VEELS) was exploited to investigate the $Ga^+$ ion damage in $Al_2O_3$, InP and InGaAs, where each target material has been shown to yield different process for altering the clustering of recoil atoms: metallic Ga, metallic In and InGaP clusters in $Al_2O_3$, InP and InGaAs respectively.Supporting simulations based on Monte Carlo and crystal orbital Hamiltonianindicate that the chemical constitution of cascade induced nano-precipitates is a result of a competition between interstitial/vacancy consumption rate and preferential bond formation.

**INTRODUCTION:**

An understanding of particle induced physical or chemical damage in compound materials has always been the subject of nuclear physics[1-5] and advanced material processing[6-11]. In the last decades, various mechanisms of ion beam irradiation damage have been predicted through density function theory[12-14] and hybrid molecular dynamics/Monte Carlo (MD/MC) simulations[15-19], including the production/recombination of Frenkel pairs[20,21], cascade induced interstitial/vacancy clustering[22,23] and long range glissile loop absorption[24], etc. These theories are of importance to describe the projectile ion dynamics in bulk materials. In fact, contributions by the chemical reactions have been totally ignored until the Hofsäss and Zhang proposed the new concept of "Surfactant sputtering"[25-27]. During ion beam sputtering, additional phase may be formed by chemical reations between incident ions and target atoms, as well as between target atoms. This will cause a fatal impact on the sputtering result, including changes in elemental sputtering rate, mass redistribution related to recoil atoms, local strain and phase separation, further altering the sputtering rate. The conventional techniques used for determining ion beam damage can be summarized as: 1) Rutherford backscattering, 2) X-ray scattering in synchrotrons, 3) atom probed tomography and 4) transmission electron microscopy. With respect to atom probed tomography and transmission electron microscopy. Rutherford backscattering and X-ray scattering have relatively poor spatial resolution, where the local fragment and atomic clustering at nanometer scale are difficult to be resolved (~30 nm)[28]. Atom probed tomography offers atomic resolution in 3D, unfortunately, the preferential ionization of certain atomic species during laser induced desorption is still a challenge for measuring the local chemical constitution[29]. Similar issues have been addressed in advanced scanning transmission electron microscopy (ATEM), where the development of aberration correctors has provided a possibility to probe chemical fluctuations at the nanometer scale, for example, in combination with energy dispersive X-ray spectroscopy (EDXS) or electron energy loss spectroscopy (EELS) and high angle annular dark field imaging (HAADF), the ion irradiation induced structural and chemical modification can be studied at near-atomic scale. However, both HAADF and EDXS have little sensitivity to recognize the chemical bonding, creating the obstacle for understanding the insight of ion/matter interaction.

Therefore, in this work, we investigate the cascade damage of $Ga^+$ ions in compound materials, where the chemical bonding in nano-precipitates formed can be evaluated. This has been possible using the monochromated valence electron energy loss (VEEL) spectroscopy that allows the exploration of individual material phase at the nanoscale. In combination with HAADF and EDXS techniques, we have shown the description of metal clustering during self-annealing should involve the understanding of atomic bonding ability. For $Ga^+$ ion implanted $Al_2O_3$, the priority of Frenkel pair recombination in $Al_2O_3$ and Al-O bond formation could lag the incorporation of Ga, leading to the precipitation of metallic Ga clusters. While for Ga implantation in InP, although the recoiled In and P atom owns a superior vacancy/interstitials short range recombination rate, the preferential Ga-P bond formation allow the yield of an InGaP ternary alloy. Finally, in case of cascade in InGaAs, the heterogeneous distribution of interstial and vacancy could give rise to the long range glissile loop consumption. The competitive In-As and Ga-As bond formation could intercept the excessive In incorporation, resulting in the precipitation of In nano-clusters.

EXPERIMENT:

A 500 nm InGaAs thin layer on 6 inches InP (001) substrate and commercial single crystalline α-$Al_2O_3$ was purchase from Epihouse Optoelectronics Co.,Ltd and Suzhou Crystal Silicon Electronic&Technology Co.,Ltd respectively. The $Ga^+$ ion damaged TEM specimens were prepared in a Thermo-Fisher Helios 600i equipped with a focused $Ga^+$ ion beam system. The 30 kV $Ga^+$ ion beam current was used to extract a 1.5 μm thick lamella from bulk sample with a beam current of ~2.5nA, the specimens were further thinned by 2 kV $Ga^+$ ion beam with beam current of 23 pA with an incident angle of 2°, which enhance the ion implantation damage with respect to conventional FIB polishing process (~9 pA). To eliminate the ion sputtering induced surface amorphization and oxide layer, surface polishing was then conducted in a Fischione Nanomill 1040, where the accelerating voltage and beam current were set as 900 eV and 150 pA, respectively. Analytical transmission electron microscopy (TEM) was carried out on a Thermo-Fisher Titan Themis G2 double aberration corrected transmission electron microscope operated at 60 kV. The microscope is equipped with a monochromator and Gatan Quantum ER965 EELS system that allows an energy resolution of ~150 meV. The energy dispersive X-ray (EDX) elemental maps were recorded from Super-X silicon drift detectors, where the energy resolution was nominally as 136 eV.

**RESULT AND DISSCUSSION：**

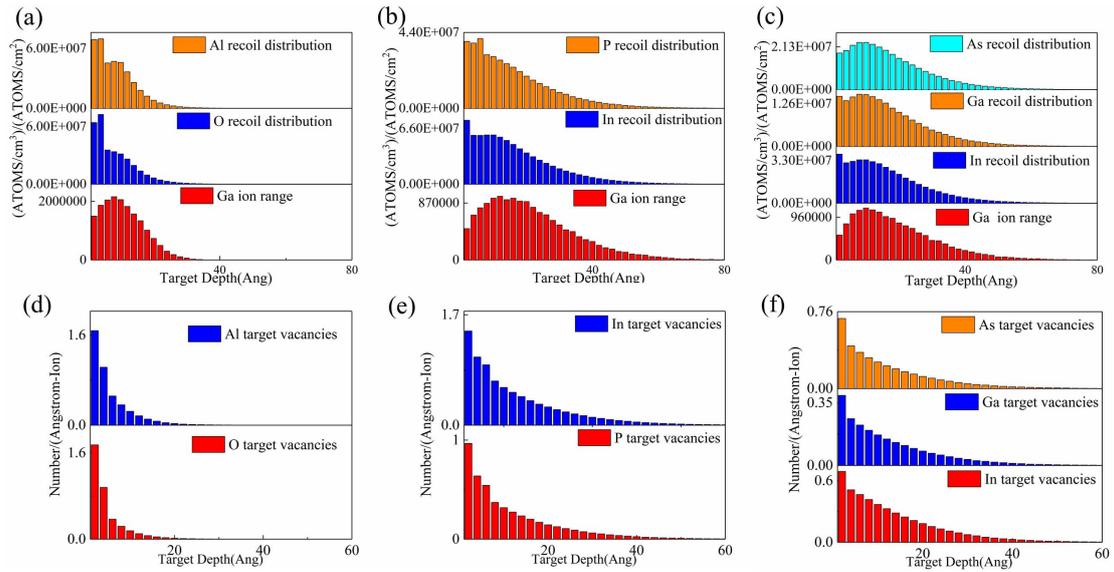

Fig.1 The recoiled atom distribution and Ga$^+$ ion range in a) Al$_2$O$_3$,b) InP, c) InGaAs. (d-f) vacancies production in d) Al$_2$O$_3$,e) InP and f) InGaAs.

The target atom recoil and vacancy distribution in **Fig. 1** were performed by Monte Carlo simulation based in Stopping and Range of Ions in Material (SRIM). The incident angle of 2 kV Ga$^+$ ion beam was set as 2º, which correlates with FIB experiment set up. For binary alloy in **Fig. 1(a,b,d,e)**, the homogenous distribution of interstitials and vacancy in Al$_2$O$_3$ and InP were observed, indicating a dominated Frenkel pairs production. Regarding to the Ga$^+$ absorbed by Al/In vacancy, since the maximum Ga$^+$ ion range was found ~30 Å underneath the sputtered surface, therefore, the heterogeneous distribution of Al/In vacancies and Ga interstitials is expected to represent the predominated damage as localized Ga clusters in Al$_2$O$_3$ and InP. On the contrast, for Ga implantation in InGaAs, a heterogeneous distributed interstitial/vacancie are observed, leading to the formation of local interstitial/vacancy clusters due to the cascade and subcascade events[30]. By exploiting the HAADF images together with EDX spectroscopy in **Fig. S1**, the Ga implantation damage in Al$_2$O$_3$, InP and InGaAs seems to be converged with prediction of Monte carlo simulation, where the Ga/In riched clusters were observed.

Unfortunately, the consumption of vacancy and interstitial during the self-annealing involves the understanding of physical processes such as Frenkel pair recombination[20,21], long-range point defect migration[24], void swelling, 、vacancy evaporation and thermal effect[31], etc, which have been generally excluded from Monte Carlo simulation. Besides, the HAADF and EDXs have poor ability to analyze chemical constitution of the nano-precipitates, the observed nanoclusters could be either electron/ion irradiation induced metallic Gallium/Indium clusters[32,33] or indicate phase separation into In poor and rich semiconductor compounds[34], left the cascade/subcascade induced interstitial/vacancy annihilation elusive.

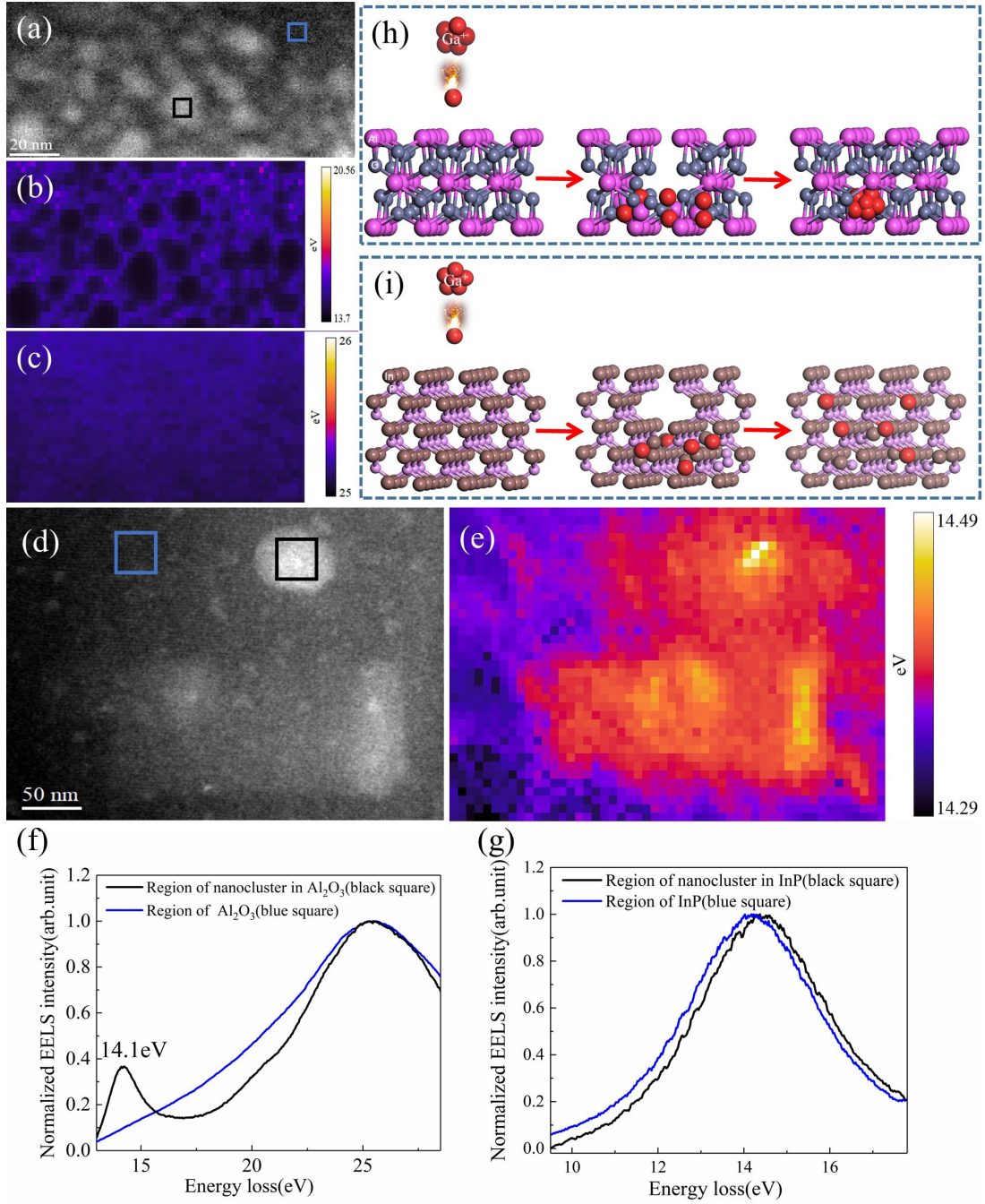

Fig.2 (a) HAADF of Ga implantation into $Al_2O_3$. (b-c) The plasmon energy maps of Ga and $Al_2O_3$. (d) HAADF of Ga implantation into InP. (e) The plasmon energy maps of InP and InGaP. (f-g) VEEL spectra of reference and ion damaged region in f) $Al_2O_3$ and g) InP. (h-i)Schematic diagram of interstitial/vacancy production and annihilation in h) $Al_2O_3$ and i) InP, respectively.

To investigate the local chemical constitution of the ion induced atomic clustering, plasmon energy loss spectroscopy was proven to be an ideal approach for analyzing local chemical constitution in ternary III-V semiconductors[29,34,35]. According to free electron approximation[36], the frequency dependent plasmon energy is given by[37]:

$$E_{P,F} = \hbar\omega_P = \hbar\sqrt{\frac{Ne^2}{V(x)m_0\varepsilon_0}} \quad (1)$$

where $E_{P,F}$ is the free electron plasmon energy in EELS spectrum, $\omega_P$ the plasmon frequency, $\hbar$ the reduced Planck constant, $N$ the number of valence electrons per unit cell, $e$ the elementary charge, $V(x)$ the volume of unit cell corresponding to doping concentration, $m_0$ the electron mass, and $\varepsilon_0$ the permittivity of free space. Many metals and narrow bandgap semiconductors (GaAs, InAs and InP etc)have sharp plasmon peaks near the value predicted by this model. For ternary InGaAs, if the In rich nanoclusters observed were due to high In content ternary alloy, the expansion of unit cell volume could lead to a decrease of valence electron density, resulting in a lower plasmon energy than its binary alloy (GaAs: 15.8 eV), while in case of precipitation of metallic In clusters in InGaAs, two plasmon peaks would be expected to emerge in VEELS due to the different numbers of valence electrons per unit cell between metallic In and InGaAs. For wide bandgap binary alloy such as $Al_2O_3$, the plasmon excitation should involve a bound oscillation $E_g$, where the $E_g$ denote the bandgap energy. the modified semi-free electron model could lead to a blue shift of plasmon peak energy towards $E_{ps}=\sqrt{E_p^2+E_g^2}$ [37]. For $Ga^+$ ion implanted $Al_2O_3$, if the Al site is occupied by a Ga atom instead, the decrease of bandgap and expansion of lattice volume could contribute to a plasmon peak shift towards lower energy (red shift). However, once the Ga rich area is dominated by metallic Ga, the wide bandgap of $Al_2O_3$ (~8.7eV)[38], the $E_{ps}$ of $Al_2O_3$ is expected to be significant larger than $E_p$ of Ga, therefore, the individual plasmon peaks of Ga and $Al_2O_3$ can be easily resolved in the spectrum, once a monochromator is used. In contrast to $Al_2O_3$, considering the Ga atoms incorporated into InP lattice, the formation of InGaP ternary alloy would result in a blue shift of InP plasmon energy, while for Ga clusters precipitated in InP, similar to metallic Ga clustering in $Al_2O_3$, two individual plasmon peaks are expected. Based on the principle of plasmon loss spectroscopy, VEEL spectrum imaging was performed to determine the chemical constitution of nanoclusters in $Al_2O_3$ (**Fig.2(f)**). In **Fig. 2(f)**, by extracting the spectrum from a single nanocluster in **Fig. 2(a)** (black square area), an additional plasmon peak was observed at 14.1 eV, which correlates well with the metallic Ga plasmon energy[39]. To exclude the existence of AlGa or $(AlGa)_2O_3$ alloy, the Ga and $Al_2O_3$ plasmon peak were modelled by multiple linear least square (MLLS) fitting via Lorentzian functions[40]. The maps of chemical constitution shown in **Fig. 2(b,c)** are obtained. As can be seen in **Fig. 2(a)** and **(c)**, the metallic Ga plasmon resonance at 14.1 eV can be observed where nanoclusters are. By observing the plasmon energy map of $Al_2O_3$ (**Fig. 2(c)**), of note, the structure of nano-precipitate has totally disappeared, referring a high purity of $Al_2O_3$ in the field of view. To explain the observation, the Frenkel pair recombination rate (short range interaction) and one dimensional glissile loop absorption rate (long range interaction) aswere adopted (eq (2))[24, 41].

$$Q=\begin{cases} g(1-e^{-kt}) & \text{(short range interaction)} \\ (3\Omega\sqrt{\pi}/4)^{\frac{2}{3}}(k_g x^{\frac{2}{3}} D_g C_g/\Omega) & \text{(long range interaction)} \end{cases} \quad (2)$$

where $g$ is the generation rate of defects, $k$ a proportionality constant, and t the time, $\Omega$ the atomic volume, $k_g$ the sink strength of the vacancy cluster. $x$ the size of the vacancy cluster, $D_g$ the diffusiion coefficient of the interstitial/vacancy, and $C_g$ the concentration of interstitial/vacancy. In case of $Al_2O_3$ irradiated by $Ga^+$ ion (**Fig. 1(a,d)**), the homogenous distribution and superior concentration of recoil atoms in $Al_2O_3$ (one order of magnitude larger than $Ga^+$ ion range in Fig. 1(a)) allows the vacancy absorption to be dominated by short range interaction, which yield a tremendous rate of Frenkel pair recombination for restricting long range occupation of Al vacancy by Ga atoms. Meanwhile, as shown in **Fig. 3(a,b)**, the integral of crystal orbital Hamilton population (icohp) of $Al_2O_3$ (39) is slightly larger than that of $Ga_2O_3$ (34.9), referring a preferential formation of Al-O bond. Therefore, due to the superior rate of interstitial/vacancy consumption and priority of A-O bond formation in $Al_2O_3$, the limited recombination of $Ga^+$ ion and Al vacancy could brings the precipitation of metallic Ga clusters. The schematic diagram of $Ga^+$ ion implantation in $Al_2O_3$ was shown in **Fig. 2(h)**.

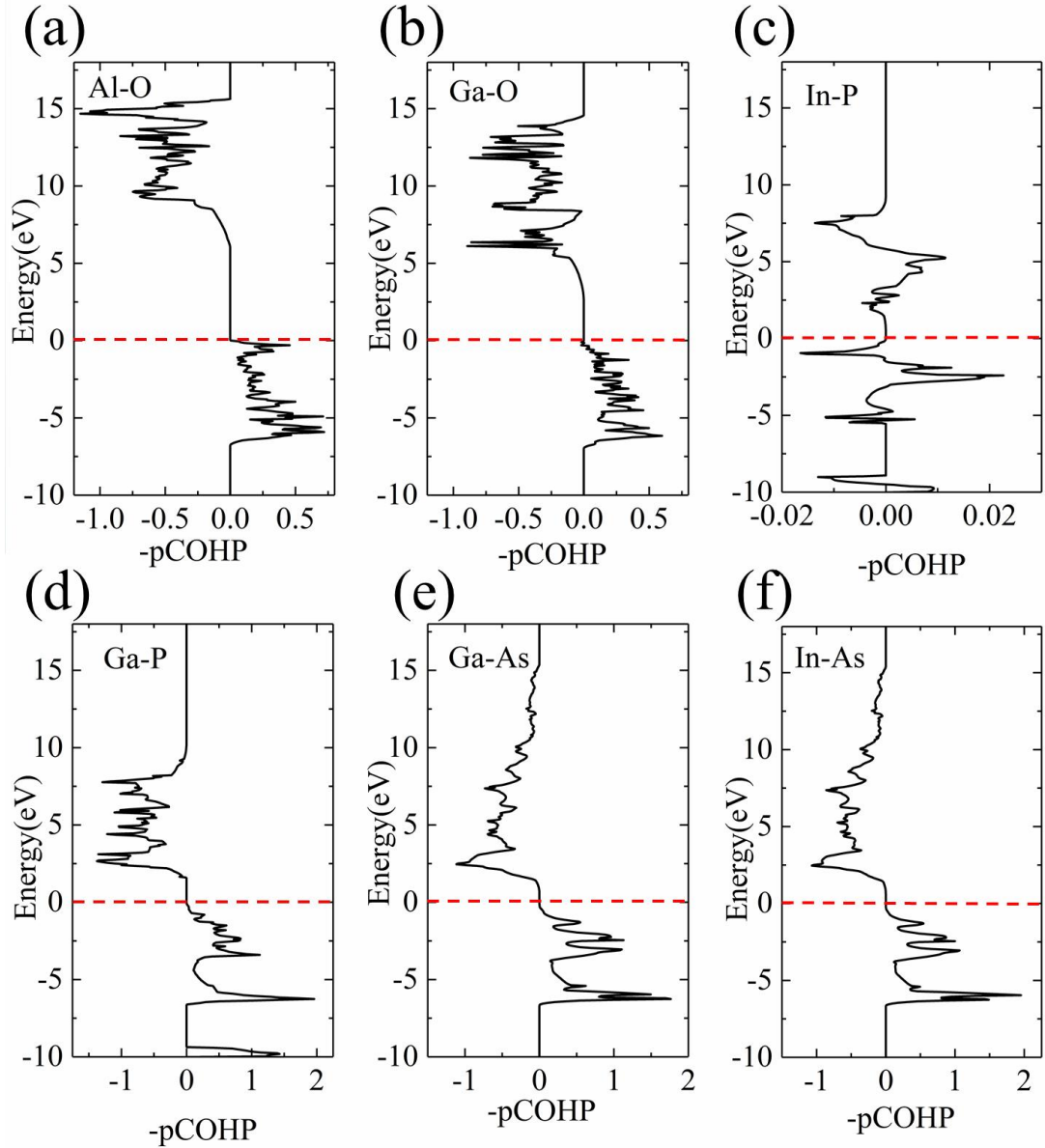

Fig.3 (a-f)Crystal orbital hamilton population(COHP) calculation of a) Al-O, b) Ga-O, c) In-P, d) Ga-P, e) Ga-As and f) In-As.

Regarding to Ga implantation in InP, compared with the InP bulk plasmon spectrum (black square region in **Fig. 2(d)**), a notable blue shift of the plasmon peak and also an absence of metallic Ga plasmon resonances was observed in the region of nanoclusters, suggesting the formation of ternary InGaP during Ga$^+$ ion implantation into InP. Similar to Al$_2$O$_3$, the InP owns advantages in short-range Frenkel pair recombination rate . Nevertheless, the privilege of Ga-P bond formation (icohp value of GaP=24.2 versus 0.1 for InP) could give rise to the effective Ga incorporation into InP lattice, resulting in the clustering of InGaP ternary alloy. The mechanism of interstitial/vacancy annihilation in InP is proposed in **Fig. 2(i).**

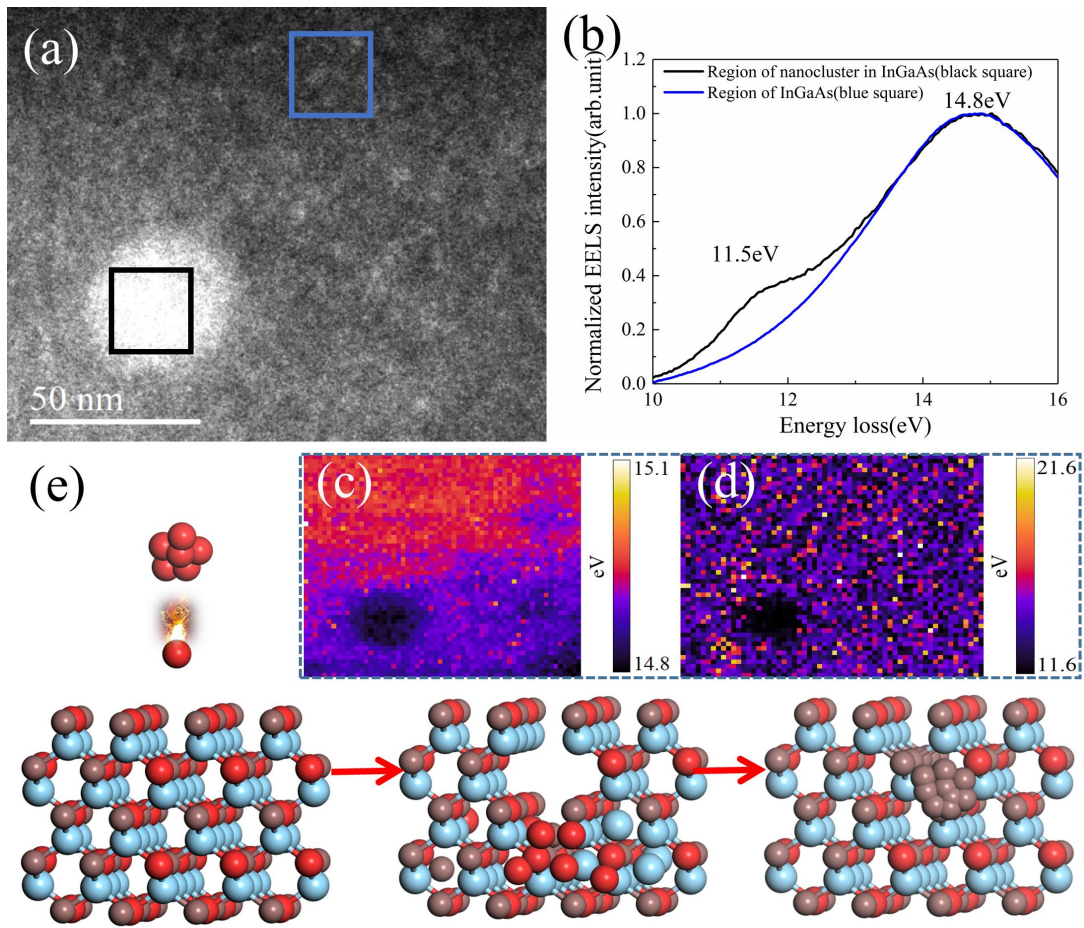

Fig. 4 (a)HAADF of Ga+ ion damaged region in InGaAs. (b) VEEL spectra of reference (blue square region) and nano-precipitates (black square region) in InGaAs. (c-d)The plasmon energy maps of InGaAs and In. (e) Schematic diagram of In segregation under Ga+ ion irradiation.

Finally, plasmon loss spectroscopy was used to analyze the chemical constitution of nano-precipitates in InGaAs. In containing semiconductor usually suffer from electron irradiation damage for beam currents above ~100 pA at 200 kV accelerating voltage[42], therefore, the optimization of probe dose rate as well as the accelerating voltage is of importance for minimizing the electron irradiation damage. In our experimental set up, the beam current of a monochromated 60 kV electron probe could approach~20 pA within a 2 nm pixel size, Rastering the InGaAs based quantum well with this convergent beam for 30 min, any electron beam damage remains invisible (supplementary data **Fig. S2**). Therefore, the optimized accelerating voltage and electron dose level ($10^8$ atoms/cm$^2$) of the probe are proven to be ideal for recording InGaAs VEEL spectrum image (**Fig. 4**). The nanoparticle in **Fig .4(a)** (region of black square) displays a shoulder peak at ~11.5 eV (**Fig. 4(b)**), which agrees with the plasmon energy of metallic In[43]. By modelling both plasmon peaks with MLLS fitting routine, the chemical constitution maps shown in **Figs. 4(c,d).** The nanoprecipitate is constituted by metallic In and high In content InGaAs ternary alloy. With respect to SRIM simulation in **Fig. 1 (c,f)**, the resultant concentration of interstitial and vacancy in InGaAs are heterogeneous distributed, where the absorption rate of interstitial/vacancy cluster should obey the long-range interaction model. By introducing the atomic volume of In (15.7 cm$^3$/mol), Ga (11.8 cm$^3$/mol) and As (13.1 cm$^3$/mol) as well as their corresponding interstitial concentration to eq (2), the In interstitials/vacancy consumption rate is approximately 2.0 and 1.4 times higher than that of Ga and As respectively, which could lead to the clustering of high In content InGaAs. Nevertheless, considering the competitive bonding ability between In-As (38.3) and Ga-As (33.2), the entire occupation of vacancies by In atoms is forbidden, left the excessive In interstitials been clustered. The schematic diagram of Ga+ ion induced In segregation in InGaAs layer is given in **Fig. 4(e)**.

## CONCLUSION:

In summary, we have proven the annihilation of interstitial/vacancy during heavy ion irradiation should involve the understanding of preferential bond formation. For $Ga^+$ ion cascades in $Al_2O_3$, the dominated Frenkel pair recombination events along with high stability of Al-O bond is responsible for producing metallic Ga clusters. For Ga sputtered InP, a preferential bond of Ga-P could prompt the Ga incorporation, enabling the clustering of an local InGaP ternary alloy. Finally, in case of $Ga^+$ ion implanted InGaAs, the competitive In-As and Ga-As bonding ability could intercept the entire incorporation of recoiled In atoms, where the aggregation of excessive In interstitials is expected to form nano-precipitates. However, due to the privilege of In interstitial concentration, the superior long range defect consumption rate could contribute to the synthesis of high In content InGaAs at nanometer scale. Our finding yields an experimental guidance for investigating the ion induced damage in compound materials, which would expect to perfect the ion/matter interaction theory on the practical side.

## SUPPLEMENTARY MATERIAL

S1: HAADF and EDX spectrum of $Al_2O_3$、InP and InGaAs, S2: HAADF、the maps of chemical constitution and EELS.

## ACKNOWLEDGEMENTS:

This work was supported by the National Natural Science Foundation of China (61975075), the Department of Science and Technology of Sichuan Province under grant numbers (2023YFH0054), the Technology and Innovation Commission of the Shenzhen Municipality (JCYJ20190809142019365), The authors acknowledge the assistance of SUSTech Core Research Facilities and the help of Dr. DongSheng He at Pico Center for the aberration corrected TEM experiments. The Strategic Cooperation Projects fostered by Zigong government and Sichuan University (No. 2021CDZG-22) and the Institutional Foundation of Institute of Chemical Materials, China Academy of Engineering Physics (No. 2021SJYBXM0217). Zhenyu Ma and Xiaoyi Wang are sincerely acknowledging the partial financial support from Dr Jinpeng Yu and Rong Wen in Sichuan Yuqian Technology Co., Ltd

## AUTHOR DECLARATIONS

**Conflict of Interest**

The authors have no conflicts to disclose.

## AUTHOR CONTRIBUTIONS:

Zhenyu Ma and Xin Zhang contributed equally to this work. Xiaoyi Wang, Yang Qiu and Thomas Walther conceived the research. Zhenyu Ma, Yong Deng, Pu Liu and Wenyu Hu conducted the TEM specimen preparation and STEM measurement. Kun Zhang, Xudong Cui and Thomas Walther advised on the modification of Ga ion implantation theory. Zhenyu Ma, Longqing Chen and Jun Zhu provided the SRIM simulation and analysis. Zhenyu Ma, Xin Zhang, Zhengshang Wang and Sen Chen conducted the COHP simulation and data processing. Zhenyu Ma, Xin Zhang, Jian Ma and Xiaoyi Wang conducted the VEELs signal processing and results analysis. All the authors contributed to data processing, results discussion and manuscript writing.

## DATA AVAILABILITY

The data that support the findings of this study are available from the corresponding author upon reasonable request.